\documentclass[twocolumn,showpacs,prb]{revtex4}
\newcommand{\figurewidth}{\columnwidth}
\usepackage{graphicx}
\usepackage{amssymb}

\begin{document}

\title{Finite size scaling of the correlation length above the upper critical
dimension}

\author{Jeff L.~Jones}
\affiliation{Department of Physics,
University of California,
Santa Cruz, California 95064}

\author{A.~P.~Young}
\homepage[Homepage: ]{http://bartok.ucsc.edu/peter}
\email[Email: ]{peter@bartok.ucsc.edu}
\affiliation{Department of Physics,
University of California,
Santa Cruz, California 95064}


\begin{abstract}
We show numerically
that correlation length at the critical point
in the five-dimensional Ising model varies
with system size $L$ as $L^{5/4}$, rather than proportional to
$L$ as in standard finite size
scaling (FSS) theory. Our results confirm a hypothesis
that FSS expressions in dimension $d$
greater than
the upper critical dimension of 4 should have $L$ replaced by $L^{d/4}$ for
cubic samples with periodic boundary conditions. We
also investigate numerically the logarithmic corrections to FSS in $d = 4$.
\end{abstract}

\maketitle

\section{Introduction}
\label{sec:introduction}

Finite size scaling\cite{barber:83,privman:90} (FSS)
has been extremely useful in extrapolating numerical
results on finite systems in the vicinity of a critical point to the
thermodynamic limit, in order to get information on critical singularities.
The basic hypothesis of FSS
is that the linear size of the system $L$ enters in the
ratio $L /\xi_\infty$ where $\xi_\infty$ is the correlation length of the
infinite system (which we will call the ``bulk'' correlation length for
convenience) and which diverges as the critical temperature $T_c$ is
approached like
\begin{equation}
\xi_\infty \approx  c_0 t^{-\nu} \, ,
\end{equation}
where 
\begin{equation}
t \equiv {T - T_c \over T_c} 
\end{equation}
measures the deviation from criticality. Here $c_0$ is a \textit{non universal}
``metric factor''\cite{privman:84} and we use the symbol $\approx$ to signify
``asymptotically equal to''.
Hence, if a
quantity $X$ diverges in the bulk like $t^{-y_x \nu}$,
the FSS form for the
behavior of $X$ is
\begin{eqnarray}
{X \over X_0} & \approx & L^{y_x} P^{\pm}
\left({L \over \xi_\infty}\right) \nonumber \\
& \approx & L^{y_x}  \widetilde{X}\left(c_1 L^{1/\nu} t \right) \, ,
\label{X}
\end{eqnarray}
where $X_0$ and $c_1$ are non-universal scale factors, and $\pm$ refers to
$t \gtrless 0$. The scaling functions $P^{\pm}$ and $\widetilde{X}$ are
\textit{universal}\cite{privman:84}. In the last expression in Eq.~(\ref{X})
we have taken
the argument of the function $P^{\pm}$
in the first expression to the power $1/\nu$,
in order that temperature appears linearly. This has the advantage that a
\textit{single} smooth function $\widetilde{X}$, applies \textit{both} above
and below $T_c$, whereas \textit{two} functions $P^{\pm}$ are needed in the
first expression in Eq.~(\ref{X}).

It is often convenient to consider \textit{dimensionless} quantities, because
these are expected to have $y_x = 0$. Two commonly studied examples are (i)
the ``Binder ratio''\cite{binder:81b},
\begin{equation}
g \equiv {1 \over 2}
\left( 3 - {\langle m^4 \rangle \over \langle m^2 \rangle^2}
\right) \approx W^{\pm}\left({L \over \xi_\infty}\right)  \approx
\widetilde{g}\left(c_1 L^{1/\nu} t\right) \, ,
\label{g}
\end{equation}
where $m$ is the order parameter, and (ii) the ratio of the correlation
length \textit{of the finite system} $\xi_L$ to the system
size\cite{nightingale:76,privman:84,kim:93}
\begin{equation}
{\xi_L \over L} \approx 
U^{\pm} \left({L \over \xi_\infty}\right)  \approx
\widetilde{\xi}\left(c_1 L^{1/\nu} t\right) \, .
\label{xiL}
\end{equation}
The definition of $\xi_L$ is not unique (though any reasonable definition
will give the same scaling form). We shall give one definition, which is often
used in numerical work, in the next section. Again, the scaling functions,
$W^{\pm}, \widetilde{g}, U^{\pm}$ and $\widetilde{\xi}$, are universal.

Note from Eqs.~(\ref{g}) and (\ref{xiL})
that, for dimensionless quantities like $g$ and $\xi_L/L$,
data for different sizes \textit{intersect at
the critical temperature}. Hence dimensionless quantities are
very convenient because they locate the critical temperature in a simple way,
from the
crossing point, without needing
to know the values of other quantities such as
exponents. 
Furthermore, since the scaling functions
$\widetilde{g}(x)$ and $\widetilde{\xi}(x)$ are
\textit{universal} the values of
$g$ and $\xi_L/L$ at the crossing point (i.e. at $T_c$) are also universal.

Finite size scaling,
as represented here by Eqs.~(\ref{X})--(\ref{xiL}), is expected
to be valid in the limit $L \to \infty, t \to 0$, with $L^{1/\nu} t$
arbitrary. Originally proposed on phenomenological grounds, a justification
for FSS was later
provided by Br\'ezin\cite{brezin:82} using renormalization group (RG)
arguments, at least for the case of systems without disorder (which is the
only case we discuss here). However, Br\'ezin\cite{brezin:82} also
noted that FSS
breaks down at the ``upper critical dimension'' $d_u = 4$. For $d > 4$ the
critical exponents are given by mean field theory, e.g. $\nu = 1/2$,
and the corresponding field
theory is a free theory (i.e. the fluctuations are Gaussian) since the
effective coupling constant vanishes at long length scales. This coupling
constant is irrelevant in the RG sense, but singularities occur when it tends
to zero, and so it cannot simply be set to its ``fixed point'' value of zero.
It is the singularities which come from this dangerous irrelevant variable
that lead to a breakdown of the FSS expressions in Eqs.~(\ref{X})--(\ref{xiL})
for $d > 4$.

Nonetheless, since the bulk behavior for $d > 4$ is trivial, one might imagine
that, in this limit, the size dependence can be also be expressed 
in fairly simple way and this turns out to be the case. As seems to be at
least implicit in much of the earlier
work\cite{brezin:82,binder:85,brezin:85,kenna:91,luijten:96} we make the
hypothesis that, 
for cubic samples with periodic boundary conditions:
\pagebreak
\begin{quotation}
\noindent
for $d > 4$, FSS formulae can still be
applied but with the system size
$L$ replaced by a \textit{larger} length\cite{luijten:pc}
\begin{equation}
\ell = A_1 L^{d/4}
\label{ell}
\end{equation}
where $A_1$ is non-universal.
\end{quotation}
We shall see that physically $\ell$ is 
\textit{the 
correlation length at the critical point}.  With this replacement,
Eqs.~(\ref{X})--(\ref{xiL}) become
(remember $d > 4$ here)
\begin{eqnarray}
{X \over X_0} & \approx & \ell^{y_x}
P^{\pm}\left({\ell \over \xi_\infty}\right) 
\approx  L^{ d y_x / 4} \widetilde{X}\left(c_2 L^{d/2} t \right) ,
\label{Xdgt4} \\
g  & \approx & W^{\pm}\left({\ell \over \xi_\infty}\right) 
\approx  \widetilde{g}\left(c_2 L^{d/2} t \right) , 
\label{gdgt4}\\
{\xi_L \over l } & \approx & U^{\pm}\left({\ell \over \xi_\infty}\right) ,
\quad \mbox{i.e.} \quad
{\xi_L \over L^{d/4} } \approx A_1\, \widetilde{\xi}\left(c_2 L^{d/2} t \right)
\, ,
\label{xiLdgt4}
\end{eqnarray}
with $c_2$ non-universal,
where we have noted that $\nu=1/2$ for $d > 4$. As before, the scaling
functions are universal, so the value of $g$ at the
crossing point at $T_c$ is universal. Furthermore, this
universal value has been calculated\cite{brezin:82,brezin:85}.  We see that,
at criticality, $\xi_L$ is of order $L^{d/4}$ which is much greater than $L$
for large sizes, a result which, at first, seems surprising. The value of
$\xi_L/L^{d/4}$ at criticality, however, is \textit{non universal} because of
the factor of $A_1$ in Eq.~(\ref{xiLdgt4}).  This factor occurs because $\ell$
has dimensions of length, and so, for Eq.~(\ref{ell}) to be dimensionally
correct, $A_1$ must be proportional to  $a^{1-d/4}$, where $a$ is a
microscopic length scale, e.g. the lattice spacing.  Quantities involving
microscopic length scales are not universal and so $A_1$ is not universal.

There has been extensive
discussion\cite{binder:85,luijten:96,blote:97,chen:98,chen:99,luijten:99}
as to whether
Eq.~(\ref{gdgt4}) applies to the five-dimensional Ising model in the limits $L
\to \infty, t \to 0$. Apparently it does\cite{luijten:96,luijten:99}, though there
appear to be several corrections to FSS which conspire to give a
``crossing'' for small sizes at a value of $g$ which differs from the
calculated universal\cite{brezin:85} value.

As noted above, a surprising feature of Eq.~(\ref{xiLdgt4}) is that the
correlation length
of the finite system at the critical point is greater than the
system size. To our knowledge there does not appear to have been any direct
verification of this prediction for $d > 4$
by numerical simulations. In this paper,
we confirm the prediction in 
Eq.~(\ref{xiLdgt4})
by Monte Carlo simulations on the five-dimensional Ising
model.  We also carry out similar simulations for the four-dimensional Ising
model, for which logarithmic corrections to standard FSS are
expected\cite{brezin:82}.

In
Sec.~\ref{sec:model} describe the model and some aspects of the simulations.
The results in five dimensions are presented in Sec.~\ref{sec:5d}, and the
results in four dimension are presented
in Sec.~\ref{sec:4d}. We summarize our results in
Sec.~\ref{sec:concl}.

\section{The Model}
\label{sec:model}

The Hamiltonian is given by
\begin{equation}
\mathcal{H} = - J\sum_{\langle i, j \rangle} S_i S_j \, ,
\end{equation}
where the Ising spins take values $S_i = \pm 1$, and the sites $i$ are on a
hypercubic lattice in $d$ dimensions of size $N = L^d$. We take $d = 4$ and
5, and apply
periodic boundary conditions. The sum is over nearest neighbor pairs of sites,
and from now on we set $J=1$.

\begin{center}
\begin{figure}
\includegraphics[width=\figurewidth]{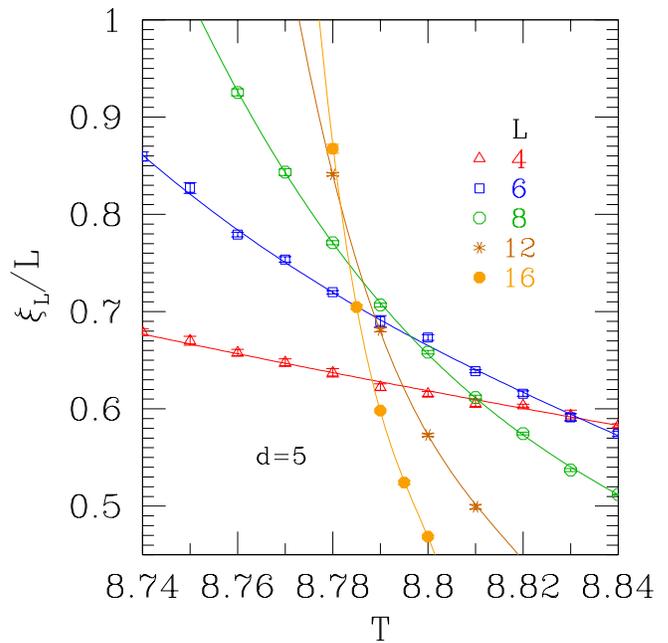}
\caption{
Data for $\xi_L/L$ in $d=5$. Clearly the data do not intersect at a common
point, as would be expected if the 
conventional FSS expression, Eq.~(\ref{xiL}), applied.
}
\label{fig:xi_overL_5d}
\end{figure}
\end{center}

\begin{center}
\begin{figure}
\includegraphics[width=\figurewidth]{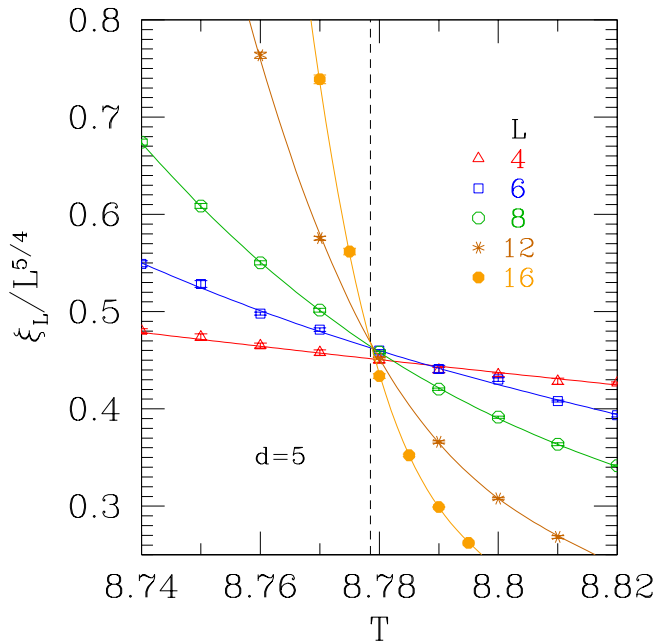}
\caption{
Data for $\xi_L/L^{5/4}$ in $d=5$. Clearly the data intersect close to
a common
point, as expected for the modified FSS expression
in Eq.~(\ref{xiLdgt4}). The
vertical line is at $T=8.7785$ which is our best estimate for $T_c$.
}
\label{fig:xi_overL54_5d}
\end{figure}
\end{center}

The magnetization per spin is given by
\begin{equation}
m = {1 \over N} \sum_{i=1}^N S_i, 
\end{equation}
and the Binder ratio is then given in terms of moments of $m$ by Eq.~(\ref{g}).
The correlation length of the finite system, is given by the following finite
difference expression\cite{kim:93}
\begin{equation}
\xi_L = {1 \over 2 \sin(k_{\rm min}/2)} \sqrt{{C(\mathbf{k}_{\rm min}) \over
C(0)} - 1} \, ,
\label{xidef}
\end{equation}
where
\begin{equation}
C(\mathbf{k}) = {1 \over N} \sum_{i, j} \langle S_i S_j \rangle \exp[i
\mathbf{k} \cdot (\mathbf{R}_i - \mathbf{R}_j)]
\end{equation}
is the Fourier transform of the spin-spin correlation function,
and $\mathbf{k}_{\rm min} = (2\pi/L) (1, 0, 0)$ is the smallest non-zero wave
vector on the lattice.
Above $T_c$ and for $L \to \infty$,
Eq.~(\ref{xidef}) 
gives the usual second moment definition of the correlation length.

We perform Monte Carlo simulations using the Wolff\cite{wolff:89} cluster
algorithm to reduce the effects of critical slowing down.

\section{Results in 5 Dimensions}
\label{sec:5d}

\begin{center}
\begin{figure}[tb!]
\includegraphics[width=\figurewidth]{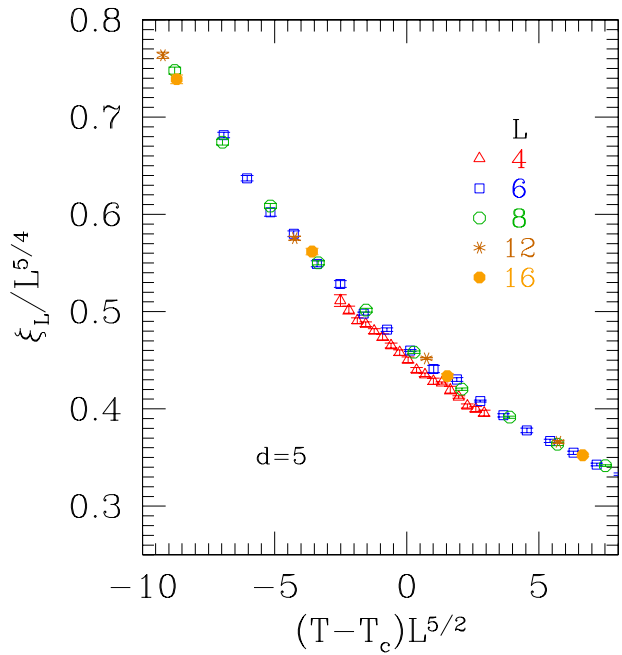}
\caption{
A scaling plot for the data in Fig.~\ref{fig:xi_overL54_5d} according to the
second expression in Eq.~(\ref{xiLdgt4}) with 
$T_c = 8.7785$.
}
\label{fig:xi_overL54_scale_5d}
\end{figure}
\end{center}

Data for $\xi_L/L$ is shown in Fig.~\ref{fig:xi_overL_5d}
for sizes $4 \le L \le 16$. According to standard FSS, Eq.~(\ref{xiL}),
the data would intersect
at a common point which is clearly not the case. However, 
according to the modified FSS
expression in Eq.~(\ref{xiLdgt4}), it is data for $\xi_L/L^{5/4}$ which should
intersect at a common point, and Fig.~\ref{fig:xi_overL54_5d}
shows that this works pretty well. Fig.~\ref{fig:xi_overL54_5d} 
therefore provides
convincing evidence that the correlation length at the critical point varies
as $L^{5/4}$ in 5 dimensions, rather than being proportional to
$L$ as would be expected in
standard FSS.

A scaling plot of the data in Fig.~\ref{fig:xi_overL54_5d} according to the
second expression in Eq.~(\ref{xiLdgt4}) is shown in
Fig.~\ref{fig:xi_overL54_scale_5d}. Note that in addition to the vertical axis
being scaled by $L^{5/4}$, rather than $L$ as in standard FSS, $T-T_c$ is
scaled by $L^{5/2}$, rather than $L^{1/\nu} \, ( = L^2)$ as in standard FSS.
Apart from
$L=4$, for which the data is consistently too low presumably because of
corrections to FSS,
the data scales well with $T_c = 8.7785$. By
considering different choices for $T_c$ we estimate that
$T_c = 8.7785(5)$, consistent with the more accurate result $8.77844(2)$ in
Ref.~[\onlinecite{luijten:99}].

\begin{center}
\begin{figure}
\includegraphics[width=\figurewidth]{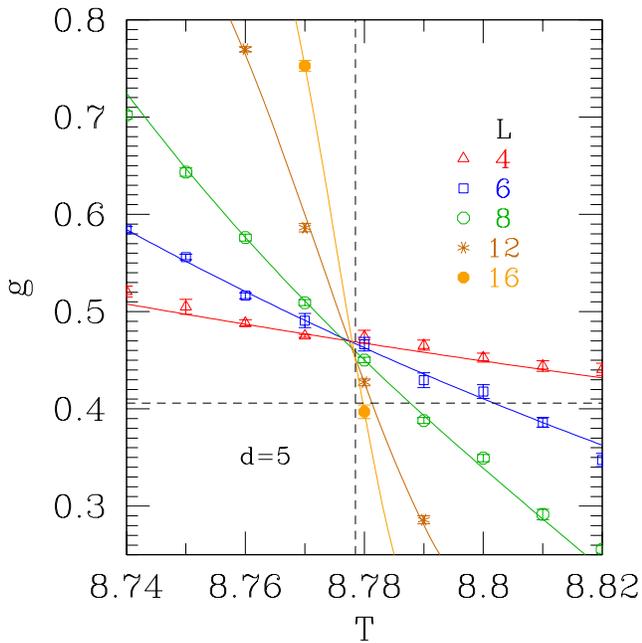}
\caption{
Data for the Binder ratio in $d=5$.
The vertical dashed line corresponds to $T = 8.7785$ which is our best
estimate of $T_c$ from the correlation length data, see
Fig.~(\ref{fig:xi_overL54_scale_5d}). The horizontal dashed line corresponds
to $g = 0.4058\cdots$, the predicted\cite{brezin:82,brezin:85} universal value.
}
\label{fig:g_5d}
\end{figure}
\end{center}

For completeness we also show results for the Binder ratio in
Fig.~\ref{fig:g_5d}.
As found in other work\cite{binder:85,blote:97,chen:98,chen:99,luijten:99},
the data
for small sizes intersect at a value of $g$ larger than the
predicted\cite{brezin:82,brezin:85}
universal value of $0.4058\cdots$.
The data for larger sizes have intersections at
somewhat smaller values and presumably\cite{luijten:99}
would reach the universal value for $L
\to \infty$.

\section{Results in 4 Dimensions}
\label{sec:4d}

\begin{center}
\begin{figure}
\includegraphics[width=\figurewidth]{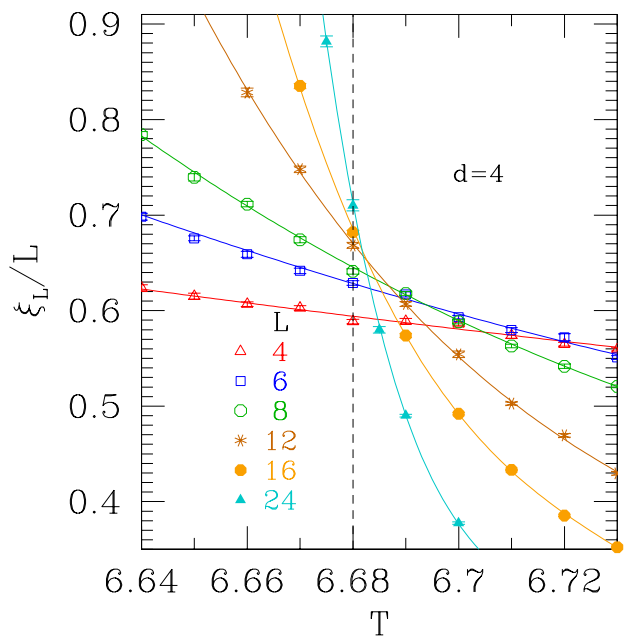}
\caption{
Data for $\xi_L/L$ in $d = 4$. According to conventional FSS, Eq.~(\ref{xiL}),
the data should have a common intersection. This is clearly not the case.
}
\label{fig:xi_overL_4d}
\end{figure}
\end{center}

\begin{center}
\begin{figure}
\includegraphics[width=\figurewidth]{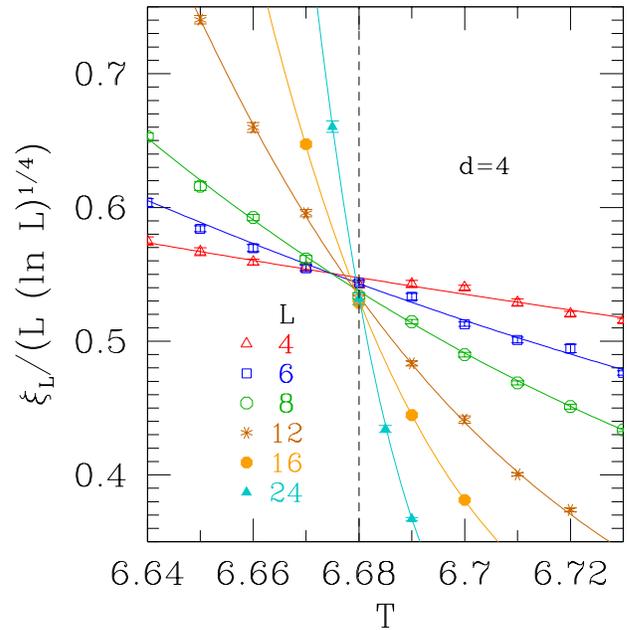}
\caption{
Data for $\xi_L/(L \ln L)^{1/4}$ in $d = 4$.  The data intersect at close to a
common point.
}
\label{fig:xi_overLlogL_4d.eps}
\end{figure}
\end{center}

In four dimensions, Br\'ezin\cite{brezin:82} argued that
$\xi_L \propto L (\log L)^{1/4}$ at criticality, and so we expect that FSS
expressions should be modified by the replacement 
\begin{equation}
L \to \ell = A_2 L (\ln L)^{1/4} \, \qquad (d = 4) \, .
\end{equation}
In
Fig.~\ref{fig:xi_overL_4d} we show a plot for $\xi_L/L$ (i.e.~without the
logarithmic factor). Clearly the data does not show a common intersection.
However, including the logarithmic factor, the plot in
Fig.~\ref{fig:xi_overLlogL_4d.eps}
shows a good intersection with only small corrections to FSS. 
The factor $\ln L$ can be replaced by $\ln (L/L_0)$ where $L_0$ is a
microscopic scale, and with an appropriate choice of $L_0$ we get sharper
intersections. However, $\ln (L/L_0) = (\ln L) (1 + \ln L_0 / \ln L)$ and so
including $L_0$ corresponds to an \textit{additive} correction to
FSS
(which vanishes only logarithmically). It is difficult to separate this from
other corrections to FSS,
and so we don't feel we can give a reliable
estimate for $L_0$. 

%
%

\section{Conclusions}
\label{sec:concl}

We have demonstrated that the FSS behavior of the correlation length (for a
cubic sample with periodic boundary conditions) in five
dimensions follows
Eq.~(\ref{xiLdgt4}), which is the expected modification of FSS for the case $d
> 4$.
This provides confirmation that the standard FSS expressions, e.g.
Eqs.~(\ref{X})--(\ref{xiL}), can be simply modified above $d=4$ by the
replacement\cite{luijten:pc} $L \to \ell \propto L^{d/4}$, which gives
Eqs.~(\ref{Xdgt4})--(\ref{xiLdgt4}).
This had been verified before for
the Binder ratio, but not, to our knowledge, for the correlation length. It is
interesting that
the correlation length at the critical point is of
order $\ell$ and hence much bigger than the system size $L$. This is
possible because the long wavelength fluctuations are non-interacting near
criticality for $d > 4$.
We also demonstrated the
expected logarithmic modification to FSS of the correlation length for $d$
precisely equal to 4.

It is also interesting to ask what are the corresponding results with $d > 4$
for other geometries and boundary conditions. For the ``strip'' geometry,
where the sample is infinite in one direction and of size $L$ in the others,
Br\'ezin\cite{brezin:82} showed that the correlation length at the critical
point varies as $L^{(d-1)/3}$ (which is reasonable since FSS is done
only with respect to the $d-1$ finite dimensions).  It is then natural to expect
that FSS will then work with $L$ replaced throughout by $L^{(d-1)/3}$.

For free boundary conditions, it seems obvious that even for $d > 4$
the behavior of the system
will be affected when $\xi_L$ becomes of order $L$, rather than only change
when $\xi_L$ becomes of order the much larger length $\ell$. Hence we expect
that the standard FSS expressions, Eqs.~(\ref{X})--(\ref{xiL}) would apply with
$\nu = 1/2$. The ratio $\xi_\infty/\ell$ may also enter but, since $\ell \gg L$ for large $L$,
such terms would presumably be corrections to the scaling terms which involve
$\xi_\infty / L$. Since FSS for models
with free boundary conditions in $d > 4$ is poorly understood, it would be
interesting to investigate such models in some detail.

\begin{acknowledgments}
We acknowledge support
from the National Science Foundation under grant DMR 0337049.  We would like
to thank Erik Luijten for helpful communications on
an earlier version of this manuscript.

\end{acknowledgments}

\bibliography{refs,comments}

\begin{thebibliography}{17}
\expandafter\ifx\csname natexlab\endcsname\relax\def\natexlab#1{#1}\fi
\expandafter\ifx\csname bibnamefont\endcsname\relax
  \def\bibnamefont#1{#1}\fi
\expandafter\ifx\csname bibfnamefont\endcsname\relax
  \def\bibfnamefont#1{#1}\fi
\expandafter\ifx\csname citenamefont\endcsname\relax
  \def\citenamefont#1{#1}\fi
\expandafter\ifx\csname url\endcsname\relax
  \def\url#1{\texttt{#1}}\fi
\expandafter\ifx\csname urlprefix\endcsname\relax\def\urlprefix{URL }\fi
\providecommand{\bibinfo}[2]{#2}
\providecommand{\eprint}[2][]{\url{#2}}

\bibitem[{\citenamefont{Barber}(1983)}]{barber:83}
\bibinfo{author}{\bibfnamefont{M.~N.} \bibnamefont{Barber}},
  \emph{\bibinfo{title}{Finite size scaling}}, in
  \emph{\bibinfo{booktitle}{Phase Transitions and Critical Phenomena}}, edited
  by \bibinfo{editor}{\bibfnamefont{C.}~\bibnamefont{Domb}} \bibnamefont{and}
  \bibinfo{editor}{\bibfnamefont{J.~L.} \bibnamefont{Lebowitz}}
  (\bibinfo{publisher}{Academic Press}, \bibinfo{year}{1983}),
  vol.~\bibinfo{volume}{8}, p. \bibinfo{pages}{146}.

\bibitem[{\citenamefont{Privman}(1990)}]{privman:90}
\bibinfo{editor}{\bibfnamefont{V.}~\bibnamefont{Privman}}, ed.,
  \emph{\bibinfo{title}{Finite Size Scaling and Numerical Simulation of
  Statistical Systems}} (\bibinfo{publisher}{World Scientific},
  \bibinfo{address}{Singapore}, \bibinfo{year}{1990}).

\bibitem[{\citenamefont{Privman and Fisher}(1984)}]{privman:84}
\bibinfo{author}{\bibfnamefont{V.}~\bibnamefont{Privman}} \bibnamefont{and}
  \bibinfo{author}{\bibfnamefont{M.~E.} \bibnamefont{Fisher}},
  \emph{\bibinfo{title}{Universal critical amplitudes in finite-size scaling}},
  \bibinfo{journal}{Phys. Rev. B} \textbf{\bibinfo{volume}{30}},
  \bibinfo{pages}{322} (\bibinfo{year}{1984}).

\bibitem[{\citenamefont{Binder}(1981)}]{binder:81b}
\bibinfo{author}{\bibfnamefont{K.}~\bibnamefont{Binder}},
  \emph{\bibinfo{title}{Finite size scaling analysis of {I}sing model block
  distribution functions}}, \bibinfo{journal}{Z. Phys. B}
  \textbf{\bibinfo{volume}{43}}, \bibinfo{pages}{119} (\bibinfo{year}{1981}).

\bibitem[{\citenamefont{Nightingale}(1976)}]{nightingale:76}
\bibinfo{author}{\bibfnamefont{M.~P.} \bibnamefont{Nightingale}},
  \emph{\bibinfo{title}{Scaling theory and finite systems}},
  \bibinfo{journal}{Physica A} \textbf{\bibinfo{volume}{83}},
  \bibinfo{pages}{561} (\bibinfo{year}{1976}).

\bibitem[{\citenamefont{{Kim}}(1993)}]{kim:93}
\bibinfo{author}{\bibfnamefont{J.~K.} \bibnamefont{{Kim}}},
  \emph{\bibinfo{title}{Application of finite size scaling to {M}onte {C}arlo
  simulations}}, \bibinfo{journal}{Phys. Rev. Lett.}
  \textbf{\bibinfo{volume}{70}}, \bibinfo{pages}{1735} (\bibinfo{year}{1993}).

\bibitem[{\citenamefont{Br\'ezin}(1982)}]{brezin:82}
\bibinfo{author}{\bibfnamefont{E.}~\bibnamefont{Br\'ezin}},
  \emph{\bibinfo{title}{An investigation of finite size scaling}},
  \bibinfo{journal}{J. Phys. (Paris)} \textbf{\bibinfo{volume}{43}},
  \bibinfo{pages}{15} (\bibinfo{year}{1982}).

\bibitem[{\citenamefont{Binder et~al.}(1985)\citenamefont{Binder, Nauenberg,
  Privman, and Young}}]{binder:85}
\bibinfo{author}{\bibfnamefont{K.}~\bibnamefont{Binder}},
  \bibinfo{author}{\bibfnamefont{M.}~\bibnamefont{Nauenberg}},
  \bibinfo{author}{\bibfnamefont{V.}~\bibnamefont{Privman}}, \bibnamefont{and}
  \bibinfo{author}{\bibfnamefont{A.~P.} \bibnamefont{Young}},
  \emph{\bibinfo{title}{Finite-size tests of hyperscaling}},
  \bibinfo{journal}{Phys. Rev. B} \textbf{\bibinfo{volume}{31}},
  \bibinfo{pages}{1498} (\bibinfo{year}{1985}).

\bibitem[{\citenamefont{Br\'ezin and Zinn-Justin}(1985)}]{brezin:85}
\bibinfo{author}{\bibfnamefont{E.}~\bibnamefont{Br\'ezin}} \bibnamefont{and}
  \bibinfo{author}{\bibfnamefont{J.}~\bibnamefont{Zinn-Justin}},
  \emph{\bibinfo{title}{Finite size effects in phase transitions}},
  \bibinfo{journal}{Nucl. Phys. B} \textbf{\bibinfo{volume}{257}},
  \bibinfo{pages}{867} (\bibinfo{year}{1985}).

\bibitem[{\citenamefont{Kenna and Lang}(1991)}]{kenna:91}
\bibinfo{author}{\bibfnamefont{R.}~\bibnamefont{Kenna}} \bibnamefont{and}
  \bibinfo{author}{\bibfnamefont{C.~B.} \bibnamefont{Lang}},
  \emph{\bibinfo{title}{Finite size scaling and the zeroes of the partition
  function in the $\varphi^4_4$ model}}, \bibinfo{journal}{Phys. Lett. B}
  \textbf{\bibinfo{volume}{264}}, \bibinfo{pages}{396} (\bibinfo{year}{1991}).

\bibitem[{\citenamefont{Luijten and Bl\"ote}(1996)}]{luijten:96}
\bibinfo{author}{\bibfnamefont{E.}~\bibnamefont{Luijten}} \bibnamefont{and}
  \bibinfo{author}{\bibfnamefont{H.~W.~J.} \bibnamefont{Bl\"ote}},
  \emph{\bibinfo{title}{Finite-size scaling and universality above the upper
  critical dimension}}, \bibinfo{journal}{Phys. Rev. Lett.}
  \textbf{\bibinfo{volume}{76}}, \bibinfo{pages}{1557} (\bibinfo{year}{1996}).

\bibitem[{lui()}]{luijten:pc}
\bibinfo{note}{Erik Luijten (private communication) has pointed out that this
  prescription needs clarification in the presence of a magnetic field $h$.
  Standard FSS scaling functions then have a second argument, $h L^\sigma$, and
  the issue is what value to take for $\sigma$ before the transformation $L \to
  \ell$ is made. If hyperscaling is satisfied then $\sigma$ is equal to both
  $y_h \equiv (d+2-\eta)/2$ and $\Delta/\nu = (\beta+\gamma)/\nu$. However, if
  hyperscaling is not satisfied, it is the latter expression which must be
  taken in order to get the correct behavior of bulk thermodynamic quantities.
  For $d > 4,$ we have $(\beta+\gamma)/\nu = 3$ and, with the replacement $L
  \to \ell$, the field enters the scaling functions in the combination $h
  L^{3d/4}$, in agreement with earlier work\cite{binder:85,luijten:96}.}

\bibitem[{\citenamefont{Bl\"ote and Luijten}(1997)}]{blote:97}
\bibinfo{author}{\bibfnamefont{H.~W.~J.} \bibnamefont{Bl\"ote}}
  \bibnamefont{and} \bibinfo{author}{\bibfnamefont{E.}~\bibnamefont{Luijten}},
  \emph{\bibinfo{title}{Universality and the five-dimensional {I}sing model}},
  \bibinfo{journal}{EuroPhys. Lett.} \textbf{\bibinfo{volume}{38}},
  \bibinfo{pages}{565} (\bibinfo{year}{1997}).

\bibitem[{\citenamefont{Chen and Dohm}(1998)}]{chen:98}
\bibinfo{author}{\bibfnamefont{X.~S.} \bibnamefont{Chen}} \bibnamefont{and}
  \bibinfo{author}{\bibfnamefont{V.}~\bibnamefont{Dohm}},
  \emph{\bibinfo{title}{Lattice $\phi^4$ theory of finite-size effects above
  the upper critical dimension}}, \bibinfo{journal}{Int. J. Mod. Phys. C}
  \textbf{\bibinfo{volume}{9}}, \bibinfo{pages}{1073} (\bibinfo{year}{1998}).

\bibitem[{\citenamefont{Chen and Dohm}(1999)}]{chen:99}
\bibinfo{author}{\bibfnamefont{X.~S.} \bibnamefont{Chen}} \bibnamefont{and}
  \bibinfo{author}{\bibfnamefont{V.}~\bibnamefont{Dohm}},
  \emph{\bibinfo{title}{Cutoff and lattice effects in the $\varphi^4$ theory of
  confimed systems}}, \bibinfo{journal}{Eur. Phys. J. B}
  \textbf{\bibinfo{volume}{7}}, \bibinfo{pages}{183} (\bibinfo{year}{1999}).

\bibitem[{\citenamefont{Luijten et~al.}(1999)\citenamefont{Luijten, Binder, and
  Bl\"ote}}]{luijten:99}
\bibinfo{author}{\bibfnamefont{E.}~\bibnamefont{Luijten}},
  \bibinfo{author}{\bibfnamefont{K.}~\bibnamefont{Binder}}, \bibnamefont{and}
  \bibinfo{author}{\bibfnamefont{H.~W.~J.} \bibnamefont{Bl\"ote}},
  \emph{\bibinfo{title}{Finite-size scaling above the upper critical dimension
  revisited: the case of the five-dimensional {I}sing model}},
  \bibinfo{journal}{Eur. Phys. J. B} \textbf{\bibinfo{volume}{9}},
  \bibinfo{pages}{289} (\bibinfo{year}{1999}).

\bibitem[{\citenamefont{Wolff}(1989)}]{wolff:89}
\bibinfo{author}{\bibfnamefont{U.}~\bibnamefont{Wolff}},
  \emph{\bibinfo{title}{Collective {M}onte {C}arlo updating for spin systems}},
  \bibinfo{journal}{Phys. Rev. Lett.} \textbf{\bibinfo{volume}{62}},
  \bibinfo{pages}{361} (\bibinfo{year}{1989}).

\end{thebibliography}

\end{document}